\documentclass{emulateapj}

\shortauthors{Brittain, S. et al.}
\shorttitle{V1647 Ori}

\usepackage{natbib}

\begin{document}

\title{Near Infrared Spectroscopic Study of V1647 Ori}

\author{Sean D. Brittain}
\affil{Department of Physics and Astronomy, Clemson University, Clemson, SC 29634-0978}
\email{sbritt@clemson.edu}

\author{Terrence W. Rettig}
\affil{Center for Astrophysics, University of Notre Dame, Notre Dame, IN 46556}

\author{Theodore Simon}
\affil{Eureka Scientific Inc., 1537 Kalaniwai Place, Honolulu HI 96821}

\author{Erika L. Gibb}
\affil{University of Missouri at St. Louis, 8001 Natural Bridge Road, St. Louis, MO, 63121}

\author{Joseph Liskowsky}
\affil{Department of Physics and Astronomy, Clemson University, Clemson, SC 29634-0978}

\begin{abstract}
We present new high-resolution infrared echelle spectra of V1647 Ori, the young star that illuminates McNeil's nebula.  From the start, V1647 Ori has been an enigmatic source that has defied classification, in some ways resembling eruptive stars of the FUor class and in other respects the EXor variables. V1647 Ori underwent an outburst in 2003 before fading back to its pre-outburst brightness in 2006. In 2008, it underwent a new outburst.  In this paper we present high-resolution $K-$band and $M-$band spectra from the W. M. Keck Observatory that were acquired during the 2008 outburst. We compare the spectra to spectra acquired during the previous outburst and quiescent phases. We find that the luminosity and full width at half maximum power of Br $\gamma$ increased as the star has brightened and decreased when the star faded indicating that these phases are driven by variations in the accretion rate.  We also show that the temperature of the CO emission has varied with the stellar accretion rate confirming suggestions from modeling of the heating mechanisms of the inner disk (e.g. Glassgold et al. 2004). Finally we find that the lowest energy blue-shifted CO absorption lines originally reported in 2007 are no longer detected. The absence of these lines confirms the short-lived nature of the outflow launched at the start of the quiescent phase in 2006.  
\end{abstract}

\keywords{accretion, accretion disks --- stars: flare --- stars: formation --- stars: individual (V1647 Ori) --- stars:winds,outflows}

\clearpage

\section{Introduction}
In November 2003 V1647 Ori began to brighten and illuminate a reflection nebula, which became known as McNeil's Nebular Object (McNeil et al. 2004). Photometry compiled by Acosta-Polido et al. (2007) shows that V1647 Ori brightened by four magnitudes in the Cousins $I_C$ band in about three months. We refer to this event as the 2003 outburst. The star faded by $\sim$ 1 magnitude over the next 600 days, then faded an additional three magnitudes in about 100 days. For a detailed summary of the 2003 outburst, see Aspin\& Reipurth (2009). In August 2008, Itagaki (2008) reported the renewed brightening of V1647 Ori, suggesting that the star had undergone another outburst. We refer to this event as the 2008 outburst. This is the third recorded outburst as archival data indicates that it brightened in 1966 (Aspin et al. 2006). 

During the onset of the 2003 outburst, the hydrogen and helium lines had P-Cygni profiles. The absorption component of these lines was shifted by several hundred kilometers per second - indicative of a hot wind with an expansion velocity of several hundred kilometers per second (Brice\~{n}o et al. 2004; Reipurth \& Aspin 2004; Vacca et al. 2004, Fedele et al. 2007).  The HI absorption features varied considerably in strength throughout 2004 and 2005 (Ojha et al. 2006; Gibb et al. 2006; Fedele et al. 2007), and these blue-shifted features have reappeared in the same lines at the outset of the 2008 outburst (Aspin et al. 2009).  

Over the course of the 2003 outburst, the infrared lines of CO also underwent significant changes.  At the start of the outburst, both the overtone bandheads (Vacca et al. 2004) and the fundamental ro-vibrational CO lines appeared in emission (Rettig et al. 2005).  Following the end of the 2003 outburst, February 2006, an absorption component with a blue shift of -35 km s$^{-1}$ appeared superimposed on the emission components in the fundamental lines. The relative strengths of these post-outburst absorption features were consistent with 700 K gas (Brittain et al. 2007).  The absorption feature was absent in observations of the v=1-0 P30 line nearly a year later, suggesting that the mechanism driving the outflow had been brief (Brittain et al. 2007). 

It is known that the truncation radius of an accretion disk scales with the stellar accretion rate (e.g. Bouvier et al. 2007 and references therein) and that rapid changes in the truncation radius can lead to outflows (Balsara 2004). With that in mind, Brittain et al. (2007) speculated that the post-outburst outflow was driven by the realignment of the stellar magnetic field as the accretion rate suddenly dropped.  Aspin et al. (2008; hereafter ABR08), on the other hand, have suggested that the fading was due to a combination of increased extinction and a more modest decrease in the accretion rate. If so, then it is possible that the truncation radius of the disk did not shift as radically as Brittain et al. (2007) supposed and that the relatively brief CO outflow has a different origin. 

In order to better characterize the evolution of the inner disk, we present high-resolution $K-$band and $M-$band spectra of V1647 Ori. We compare the evolution of Br $\gamma$, the v=2-0 bandhead of CO, and the fundamental CO lines spanning the years from 2004 to 2009. We also compare the entire $K-$band spectrum during the 2008 outburst to the period immediately preceding the outburst. Finally we discuss the evolution of the line profiles, gas temperature, and truncation radius of the inner disk. 

\section{Observations \& Data Reduction}
We acquired high-resolution spectra of V1647 Ori on February 18, 2006, January 31, 2009,  and February 1, 2009, with the NIRSPEC instrument at the W. M. Keck Observatory on Mauna Kea, Hawaii. NIRSPEC is a high-resolution ($\lambda/\delta\lambda\sim25,000$), near-infrared ($1-5\micron$), cross-dispersed spectrometer (McLean et al. 1998). Roughly 25\% of the $M-$band and 50\% of the $K-$band can be covered per setting. The wavelength coverage of our settings is presented in Table 1. The $K-$band spectrum acquired in 2006 was taken under poor seeing conditions ($\sim$1.2$\arcsec$) and high wind. Further, we were required to close down due to fog before a telluric standard could be acquired. The weather conditions during our 2009 observations were much better. The seeing was $\sim$0$\arcsec$.4. The telluric correction of the data taken in 2009 was achieved by observing the standard star HR~3888 at a similar airmass as V1647 Ori. Regions of the spectrum where the atmospheric transmittance was less than 50\% have been omitted. All data were acquired, reduced, and calibrated using standard observing techniques, which are described in Brittain (2004). One exception is our reduction of the spectral order containing Br $\gamma$. The standard star revealed a weak, broad photospheric feature. This order was corrected using the Spectral Synthesis Program (Kunde \& McGuire 1974), which accesses the 2008HITRAN database (Rothman et al. 2009).

\section{Results}
\subsection{$K-$band Spectrum}
\subsubsection{Metal Lines}
Aspin et al. (2009; hereafter AGR09) demonstrated the close resemblance between the $K-$band spectrum of V1647 Ori in quiescence to those of FUors and the surprisingly poor correspondence with the spectra of  EXor stars.  Similarly, they showed the correspondence with reference spectra of various normal late-K and early-M type stars of low surface gravity was poor, even after the standard star spectra were velocity broadened to 120 km s$^{-1}$ to match the FWHM of the features observed in V1647 Ori.  

In Figure 1, we compare the same spectral region acquired during the 2008 outburst to the spectra presented by AGR09.  With the exception of the Br $\gamma$ line (\S 3.1.2) and the v=2-0 CO bandhead (\S 3.1.3), the changes in the $K-$band spectrum from the quiescent phase to the 2008 outburst are quite subtle (Figure 1). The features scattered throughout the spectra appear heavily veiled but are otherwise unchanged. If these lines originate in the circumstellar disk (as has been shown for FUors; Hartmann et al. 2004), then one might expect them to become rotationally broadened as the accretion rate increases and the truncation radius moves inward, closer to the star.  Further, as a larger fraction of the inner disk participates in this absorption spectrum, one might also expect that the lines would appear less veiled. This does not appear to be the case, thus we conclude the lines are photospheric in origin. 

Assuming the luminosity of the star is constant, the normalized quiescent spectrum, $L_Q$, is related to the normalized outburst spectrum, $L_O$, by the relation

\begin{displaymath}
L_Q=\frac{L_O-r}{1-r}
\end{displaymath}

\noindent where $r$ is the intrinsic luminosity of the excess IR emission scaled to the luminosity of the system. By scaling the $K-$band spectra acquired during the 2008 outburst to the $K-$band spectra acquired during the quiescent phase (AGR09), we find that $r=0.50^{+0.05}_{-0.10}$. Thus the luminosity of the veiling material is 

\begin{displaymath}
L_v=\frac{r}{1-r}L_\star = 1.0^{+0.2}_{-0.3}L_\star, 
\end{displaymath}

\noindent indicating that the star+disk brightened by $0.8^{+0.2}_{-0.3}$ magnitudes in the $K-$band.

\subsubsection{Br $\gamma$}
Figure 2 shows that the Br $\gamma$ line shape has evolved considerably during the alternating outbursting and quiescent phases. The 2004 and 2005 spectra were acquired with SPEX at the IRTF with resolutions of $\lambda/\Delta\lambda$=2,000 (Vacca et al. 2004) and 1,200 (Gibb et al. 2006). The 2006 and 2009 spectra where acquired using NIRSPEC at the Keck telescope with a resolving power of $\lambda/\Delta\lambda$=25,000. The 2006 spectrum was smoothed with a bin of 3-pixels. The 2007 and 2008 spectra were also acquired with NIRSPEC at the Keck telescope but with a resolving power of $\lambda/\Delta\lambda$=18,000 (AGR09).  At the onset of the 2003 outburst, the full width at the half-maximum power (FWHM) of the line was $\sim$550 km s$^{-1}$.  During the quiescent phase (2006-2008), the FWHM was $\sim$150 km s$^{-1}$.  During the onset of the 2008 outburst, the line width increased to 665 km s$^{-1}$ (Aspin et al. 2009; not plotted in Fig. 2) and by January 2009, it decreased to $\sim$400 km s$^{-1}$ (Table 2; Figure 2). To our knowledge, such a correlation between the width of Br $\gamma$ and the accretion rate has not been observed before, nor has it ever been noted as a  prediction of the published disk models.  


Because Br $\gamma$ is a proxy for the stellar accretion rate of young stars and there are a large number of Br $\gamma$ measurements spanning 2004-2009 in the literature, we use these measurements to provide a measure of the stellar accretion rate of V1647 Ori. The stellar accretion rate is thought to drive the brightening of both EXor and FUor outbursts. Additionally, accretion in the disk is likely an important heating mechanism (Glassgold et al. 2004). By comparing the accretion rate to the change in CO line luminosity as star passes between outbursting and quiescent phases, we can examine the degree to which accretion heating impacts the excitation of CO. 

Muzerolle et al. (1998) found that the luminosity of the Br $\gamma$ line of the classical T Tauri stars is related to the stellar accretion rate such that,

\begin{equation}
\label{1}
\log (\frac{L_{acc}}{L_{\sun}})=(1.26\pm0.19)\log(\frac{L_{Br\gamma}}{L_{\sun}})+(4.43\pm0.79)
\end{equation}

\noindent and

\begin{equation}
\label{2}
\dot{M}=L_{acc}R_{\star}/GM_{\star}. 
\end{equation}

\noindent As the accretion luminosity increases, the luminosity of Br $\gamma$ increases because the total surface area of the funnel flows from the disk to the star increases even as the truncation radius moves closer to the star (Muzerolle, Calvet, \& Hartmann 2001). This relationship does not apply to FUors as Br $\gamma$ is generally observed in absorption (AGR09). Furthermore, in the case of the FUors, the magnetospheric region in which the Br $\gamma$ line is thought to originate is expected to be crushed under the enormous accretion rate (Muzerolle et al. 1998). However, unlike FUors, the Br $\gamma$ line of V1647 Ori has always been observed in emission. The peak accretion rate of this star during the 2003 outburst as inferred from the bolometric luminosity was determined to be $\sim$10$^{-5}$ M$_{\sun}$ yr$^{-1}$ (Muzerolle et al. 2005), which is an order of magnitude less than the typical accretion rate of FUors (Hartmann \& Kenyon 1996).  Because of the lower accretion rate, it is reasonable to suppose that the accretion geometry of V1647 Ori may be much more similar to that of classical T Tauri stars. Thus, even though the $K-$band spectrum of V1647 Ori appears to correlate better with the $K-$band spectra of FUors than classical T Tauri stars, we adopt the relationship between the Br $\gamma$ line luminosity and stellar accretion rate to follow the evolution of the accretion rate of V1647 Ori from the 2003 outburst into quiescence and back into outburst. To do so, we compile a collection of Br $\gamma$ line luminosities from the 2003 outburst through the return to quiescence and into the onset of the 2008 outburst. 

Several of the Br $\gamma$ observations in the literature only report the equivalent width of the line. Since the brightness of V1647 Ori has varied by 3 magnitudes in the $K-$band, we have collected photometric observations bracketing the time the spectra were taken. The equivalent width of the emission line is then scaled to the continuum flux level to obtain the line flux. The variation of the reported $K-$band magnitude during the dates surrounding the observation is adopted as the uncertainty in the line flux. The fluxes are converted to luminosities by adopting a distance of 450~pc and are presented in Table 2. 

Several authors have reported the accretion rate of V1647 Ori based on the relationship between the luminosity Br $\gamma$ and the stellar accretion rate but using different values for the distance, stellar radius, stellar mass, and $K-$band extinction. Because these parameters are poorly constrained (and thus vary from paper to paper), we have recalculated the accretion rate using a consistent set of assumptions. To calculate the accretion rate over the dates spanning 2004-2009, we adopt M$_\star$=0.8M$_\sun$, R$_\star$=5R$_\sun$,  A$\rm_V$=9 (A$\rm_K$=1.0) for the observations taken during the outbursts, and A$\rm_V$=19 (A$\rm_K$=2.1) for the quiescent observations  following ABR08. Using equations 1 and 2, we use these Br $\gamma$ luminosities to calculate the stellar accretion rate and present these rates in the final column of Table 2. The accretion rate dropped by a factor of 22 from its maximum during the 2003 outburst to its minimum during the quiescent phase.

\subsubsection{Overtone CO lines}
In contrast to the observations during the 2003 outburst and the quiescent phase, we did not detect the v=2-0 bandhead of CO in either emission or absorption in January 2009 during the 2008 outburst (Fig. 1a). The absence of CO bandhead emission is unusual for outbursting EXors but common for T Tauri stars. As recently as January 26, 2008, the depth of the overtone bandhead of CO was $\sim$25\% (AGR09). In our data, the spectrum is flat in the region where the bandhead is expected, but there is a notable correspondence between the other features in that spectral region (\S 3.1). This suggests that the CO v=2-0 bandhead is not simply veiled by the increased $K-$band continuum, but rather that the photospheric CO v=2-0 bandhead absorption feature is filled-in by circumstellar CO v=2-0 bandhead emission. In principle, one could model the CO overtone bandhead observed during quiesence, correct it for veiling, then substract it from the spectrum acquired during the 2008 outburst to infer the overtone emission. However, it is more straightforward to model the fundamental spectrum which is less affected by the stellar photosphere.  Since the v=2-1 emission lines originate in the same state as the v=2-0 emission lines, they probe the same gas as the overtone emission lines. 

\subsection{$M-$band Spectra}
Figure 3 shows the $M-$band spectrum of V1647 Ori acquired during the 2008 outburst. The fundamental ro-vibrational CO spectrum is quite complicated.  The emission lines are due to $\Delta$v=1 CO transitions. Superimposed on the low-J emission lines (J$^\prime<$4) are narrow absorption lines from cold interstellar gas. The broad absorption line centered at 2139.5 cm$^{-1}$ is due to CO ice.  Since 2004, the narrow absorption lines and CO ice feature have not varied beyond the signal to noise of our spectra ($\lesssim$3\%). This is consistent with our previous suggestion that this material lies in the foreground and is unaffected by the outbursts of V1647 Ori (Rettig et al. 2005; Gibb et al. 2006). Conspicuously absent is the ubiquitous Pf $\beta$ emission line at 2149 cm$^{-1}$. This dominating emission feature is always present in CO emission spectra of T Tauri and Herbig Ae/Be stars (\emph{c.f.} Najita et al. 2003; Blake \& Boogert 2004; Brittain et al. 2007). It is not generally observed in spectra of FUors (Brittain unpublished data). Its absence provides further evidence that V1647 Ori is not simply a vigorously accreting classical T Tauri star. 

To facilitate our analysis of the circumstellar CO lines, we have removed these foreground features by calculating the absorption spectra using the parameters presented in Rettig et al. (2005) and Gibb et al. (2006). The corrected spectra are presented in figure 4. The v=2-1 emission lines are broader than the high-J emission lines which are broader than the low-J and mid-J emission lines as is expected for gas in a Keplerian disk with a negative temperature gradient. To determine the gas distribution in the disk, we have synthesized a disk spectrum adopting the formalism described in Najita et al. (1996). We assume that the gas originates in an azimuthally symmetric Keplerian disk inclined by 60$\degr$. The gas temperature is described by a power law where $T(R)=1800(R/1AU)^{-0.35}$. We achieve our best fit for gas that extends from 0.1-12AU (Fig. 4). The inclination of this source is not well constrained, and the distribution of the gas that we infer from our modeling scales as $sin^2(i)$. If the disk is more face-on, then the gas distribution will be more compact. The effective temperature of the emission is 1830 K. This is considerably cooler than the temperature of the gas observed during the onset of the 2003 outburst ($\sim$2500K; Rettig et al. 2005), but warmer than the temperature measured during quiescence (1400K; Brittain et al. 2007). 

In addition to being warmer than the CO observed in emission during the quiescent phase, the blue-shifted ro-vibrational CO absorption lines observed 17 February, 2006 are no longer present (Fig. 5). Previously, the v=1-0 P30 blue-shifted absorption line was shown to vanish by January 2007.  By that date during the quiescent phase, the outflowing CO had either cooled or been destroyed as the material expanded and became more diffuse. Three years following the outflow, it is clear that there is no residual cool outflowing CO indicating that the engine driving the outflow was indeed short lived and the launched material was dissociated as it grew more diffuse. 

The FWHM of the v=2-1 emission lines observed during the 2008 outburst are narrower than their counterparts observed during the 2003 outburst (Fig. 6). However, the HWZI of the average v=2-1 CO emission line profile acquired during the 2003 outburst is the same to within our signal-to-noise $-60\pm5$ km s$^{-1}$. Since the truncation radius of the gas disk scales with the accretion rate, this suggests that the accretion rate during the 2003 and 2008 outbursts are similar as is indicated by the luminosity of the Br $\gamma$ emission. 

\section{Discussion}
ABR08 suggest that the fading of V1647 Ori is due at least in part to a dramatic increase in the extinction rather than simply the intrinsic fading of the central source. To explore this suggestion, we compare high resolution spectra acquired through the outbursting and quiescent phases.  During the 2008 outburst, the apparent $K-$band magnitude of V1647 Ori increased from 9.8 magnitudes in September 2006 (Acosta-Polido et al. 2007) to 7.5 magnitudes in September 2008 - comparable to the brightness achieved during the 2003 outburst.  No $K-$band photometry has been reported closer to the time when we acquired our spectrum (January 2009), so the $K-$band magnitude at the time of our observation is not entirely clear. We can estimate the relative flux of the 2009 and 2006 observations by comparing the signal to noise of the spectra. The signal-to-noise of our spectrum acquired in 2009 improved by about an order of magnitude relative to our $K-$band spectrum acquired in 2006. The integration time of the observation in 2009 was 12 minutes rather than 4 minutes and the seeing was 0$\arcsec$.4 rather than 1$\arcsec$.2. Correcting for slit loss and integration time and assuming the star was properly centered in each observation, the relative signal to noise of the spectra indicates that the star brightened by about an order of magnitude (or $\sim$2.5 magnitudes) between the 2006 and 2009 observations. This coarse estimate agrees reasonably well with the measured 2.3 magnitudes of brightening and indicates that the star did not dramatically fade between the September 2008 observation and our observation in January 2009.    

Our analysis of the $K-$band veiling indicates that intrinsic brightening of the source was  $0.8^{+0.2}_{-0.3}$ magnitudes. This leaves $1.5^{+0.2}_{-0.3}$ magnitudes of brightening unaccounted. If this additional brightening was solely the result of decreased extinction along the line of sight, this implies that $\rm A_V$ decreased by 14$^{+2}_{-3}$ magnitudes. If $\rm A_V$=9 magnitudes during the outbursts, this implies that the extinction during the quiescent phase was 23$^{+2}_{-3}$ magnitudes consistent with the extinction of $\rm A_V$=19$\pm$2 magnitudes inferred by ABR08. 

The column density of the CO outflow observed following the 2003 outburst was N(CO)=3$\times$10$^{18}$ cm$^{-2}$. If we assume that the composition of the wind and reddening law follow canonical interstellar values (i.e., N(H$_2$)/N(CO)$=1.5\times 10^{-4}$ and N(H$_2$)/A$\rm_V$=9.3$\times$10$^{20}$ ~molecules~magnitude$^{-1}$) then this column density of CO corresponds to A$\rm _V$=22 magnitudes. This supports the hypothesis that the rapid fading of V1647 Ori was due to a combination of a drop in the accretion rate and increase in the extinction. 

The increased veiling of the photospheric lines in the $K-$band cannot account for the disappearance of the v=2-0 CO bandhead during the 2008 outburst. This result indicates the presence of a circumstellar emission component. Emission by the overtone CO bandheads requires remarkably dense and hot gas over an extended region (Scoville et al. 1980; Najita et al. 1996). The fact that the CO v=2-0 bandhead absorption feature has been filled suggests that the temperature and density of the inner disk has significantly increased relative to the disk conditions in 2007 and 2008 when the CO bandheads were observed in absorption (AGR09). While this emission has increased, it has not increased to the rate seen during the 2003 outburst when the overtone CO lines were seen in emission (Vacca et al. 2004). This indicates the rebrightening is indeed due to reinvigorated accretion as suggested by our analysis of the luminosity of Br $\gamma$, though the rate has not increased to its level during the 2003 outburst.

The FWHM of the Br $\gamma$ line changed with the stellar accretion rate suggesting that the origin of the  line was indeed related to accretion. However, the accretion rate inferred from Br $\gamma$ is not consistent with Pa $\beta$ (ABR08). Why would the value for accretion inferred from these lines diverge? A more detailed study of these lines is necessary to draw conclusions on how they arise in the accretion flow. High spatial and spectral resolution study of these lines on nearby sources is crucial for testing calculations of line formation in winds and funnel flows. We do not have data on Pa $\beta$, so we cannot say whether shape scales with the brightness of the star. 

There is a systematic uncertainty in the stellar mass (25\%), stellar radius (20\%) and distance (ABR08), thus the relative accretion rate is more useful for tracking the change in the accretion rate from during the recent outbursting and quiescent phases of V1647 Ori. We have recalculated the accretion rate inferred from the Br $\gamma$ luminosity making a consistent set of assumptions (\S3.1.2) and found that the accretion rate dropped by a factor of 22 from its high during the 2003 outburst to its low during quiescence and increased by a factor of 16 during our observations during the 2008 outburst.  The truncation radius of the disk, $R_T$, is inversely and nonlinearly proportional to the accretion rate, $\dot{M}$, and given by $ R_T \propto \dot{M}^{-2/7}$ (e.g. Bouvier et al. 2007).  Thus the factor of 22 drop in the stellar accretion rate experienced by V1647 Ori would have resulted in the truncation radius being shifted outward by a factor of $\sim$2. The re-organization of the stellar magnetic field in response to such a shift may have resulted in the CO outflow observed in 2006 (Brittain et al. 2007; Romanova 2009)

Rettig et al. (2005) note that the narrow, cold fundamental CO absorption lines and CO ice feature are consistent with cold foreground material. Indeed, those features have remained unchanged through the last five years (Gibb et al. 2006, Brittain et al. 2007, Aspin et al. 2009, \S3.2) while the lines formed in the vicinity of the star have varied dramatically. The fundamental CO emission lines have largely returned to their appearance first observed during the 2003 outburst. These lines are generally symmetric and consistent with an origin from a Keplerian disk. The blue-shifted absorption lines observed in 2006 are no longer observed among the high-J lines. This was the case in the low S/N 2007 data presented by Brittain et al. (2007). There is also no evidence of a blue-shifted component among the low-J lines. Due to the sparse sampling of V1647 Ori in the $M-$band, it is not clear when the outflow began and ended nor how the temperature of the gas evolved as it expanded. Synoptic observation of the stellar accretion rate, disk truncation radius (as measured by fundamental CO emission) and various measures of the outflow will provide important insight to how the star-disk interaction evolves with the stellar accretion rate. 

\section{Conclusion}
We find that the veiling of the photospheric lines in the $K-$band increased modestly but were otherwise unchanged. Thus the underlying reason for the strong correlation between these lines and those in FUors is independent of the accretion state of the system. Further, the accretion rate of the system during the 2008 outburst is a factor of 16 greater than the smallest accretion rate observed during the quiescent phase and nearly as high as the peak accretion rate observed during the 2003 outburst. The width of Br $\gamma$ has varied with the accretion rate.  The fundamental CO spectrum indicates that the mechanism that gave rise to the blue-shifted CO lines observed as the star returned to the quiescent phase was indeed short-lived. There is no evidence of remnant cool blue-shifted gas nor has the column of cold CO absorption increased indicating that the column density of intervening material has not changed appreciably. 

\acknowledgements
The data presented herein were obtained [in part] at the W.M. Keck Observatory, which is operated as a scientific partnership among the California Institute of Technology, the University of California and the National Aeronautics and Space Administration. The Observatory was made possible by the generous financial support of the W.M. Keck Foundation. S.D.B. acknowledges support for this work from the National Science Foundation under grant number AST-0708899 and NASA Origins of Solar Systems under grant number NNX08AH90G.

\clearpage

\begin{deluxetable}{lcccccc}
\tablecaption{Log of Observations}

\tablehead{ \colhead{Date} & \colhead{Star}  & \colhead{Filter} & \colhead{Setting\tablenotemark{a}} &
\colhead{Spectral Grasp}  &  \colhead{Integration time} & \colhead{Signal to Noise} \\
\colhead{} & \colhead{} & \colhead{} & \colhead{} & \colhead{cm$^{-1}$} & \colhead{minutes} & \colhead{}}
\startdata
February 18, 2006 & V1647 Ori & K & 62.7/35.59 & 4570-4640 & 4 & 15 \\

January 31, 2009	&	V1647 Ori	&	MW	&	61.12/37.05	&	1995-2028	& 12 &	100	\\
	&		&		&		&	2129-2162	& &	60	\\
	&	V1647 Ori	&	K	&	61.93/35.49	&	4338-4403	& 8 &	90	\\
	&		&		&		&	4470-4537	& &	130	\\
	&		&		&		&	4600-4675	& &	120	\\
	&		&		&		&	4732-4792	& &	130	\\
	&	HR 3888	&	MW	&	61.12/37.05	&	1997-2028	& 4 &	140	\\
	&		&		&		&	2129-2162	& &	90	\\
	&	HR 3888	&	K	&	61.93/35.49	&	4338-4403	& 8 &	150	\\
	&		&		&		&	4470-4537	& &	200	\\
	&		&		&		&	4600-4675	& &	210	\\
	&		&		&		&	4732-4792	& &	210	\\
February 1, 2009	&	V1647 Ori	&	MW	&	60.58/37.05	&	1968-1997	& 8 &	60	\\
	&		&		&		&	2098-2129	& &	35	\\
	&	HR 3888	&	MW	&	60.58/37.05	&	1968-1997	& 4 &	100	\\
	&		&		&		&	2098-2129	& &	60	\\
\enddata
\tablenotetext{a}{Echelle and Cross Disperser settings.}
\end{deluxetable}
\clearpage

\begin{deluxetable}{lcccccccc}

\tablecaption{Br $\gamma$ Data}

\tablehead{ \colhead{Date} & \colhead{Activity Phase} & \colhead{Equivalent Width}  & \colhead{FWHM} & \colhead{Flux} & \colhead{A$\rm_K$ \tablenotemark{a}} & \colhead{L$_{Br \gamma}$} & \colhead{L$_{acc}$} &  \colhead{Accretion rate}  \\
\colhead{} & \colhead{} & \colhead{\AA} & \colhead{km s$^{-1}$} & \colhead{10$^{-17}$ W m$^{-2}$} & \colhead{mag.} & \colhead{10$^{-3}$ L$_{\sun}$} & \colhead{L$_\sun$} &  \colhead{10$^{-6}$ M$_\sun$ yr$^{-1}$}  }

\startdata
2004 Mar 08	& O & 	-5.2\tablenotemark{b}	&	\nodata				&	25.3	&	1.0	&	4.0	&	26	&	5.0	\\
2004 Mar 09	& O &	-4.6\tablenotemark{c}	&	550\tablenotemark{c}	&	26.9	&	1.0	&	4.3	&	28	&	5.4	\\
2004 Nov 04	& O &	-4.8\tablenotemark{b}	&	\nodata				&	23.8	&	1.0	&	3.8	&	24	&	4.6	\\
2004 Nov 30	& O &	-5.06\tablenotemark{d} 	&	\nodata				&	17.2	&	1.0	&	2.7	&	16	&	3.1	\\
2005 Mar 02	& O &	-5.2\tablenotemark{d}	&	270\tablenotemark{d}	&	15.5	&	1.0	&	2.5	&	14	&	2.7	\\
2006 Feb 18	& Q &	-3.1\tablenotemark{e}	&	150\tablenotemark{e}	&	1.6\tablenotemark{h}	&	2.1	&	0.7	&	2.8	&	0.55	\\
2006 May 05	& Q &	-6.9\tablenotemark{b}	&	\nodata				&	3.6	&	2.1	&	1.5	&	7.7	&	1.5	\\
2006 Sep 09	& Q &	-6.7\tablenotemark{b}	&	\nodata				&	3.5	&	2.1	&	1.5	&	7.4	&	1.5	\\
2007 Feb 22	& Q &	-5\tablenotemark{a}		&	150\tablenotemark{a}	&	2.3	&	2.1	&	1.0	&	4.4	&	0.85	\\
2007 Aug 26	& Q &	-3.9\tablenotemark{f}	&	150\tablenotemark{f}	&	1.9\tablenotemark{i}	&	2.1	&	0.8	&	3.6	&	0.70	\\
2008 Jan 26	& Q &	-1.5\tablenotemark{f}	&	150\tablenotemark{f}	&	0.9\tablenotemark{i}	&	2.1	&	0.4	&	1.3	&	0.25	\\
2008 Aug 31	& O &	-5\tablenotemark{g}		&	665\tablenotemark{g}	&	21.5	&	1.0	&	3.4	&	21	&	4.1	\\
2009 Jan 31	& O &	-3\tablenotemark{e}		&	400\tablenotemark{e}	&	12.9\tablenotemark{j}	&	1.0	&	2.0	&	11	&	2.2	\\

\enddata
\tablenotetext{a}{ABR08}
\tablenotetext{b}{Acosta-Pulido et al. 2007}
\tablenotetext{c}{Vacca et al. 2004}
\tablenotetext{d}{Gibb et al. 2006}
\tablenotetext{e}{This work}
\tablenotetext{f}{AGR09}
\tablenotetext{g}{Aspin et al. 2009}
\tablenotetext{h}{Adopt $m_K$ from Acosta-Pulido et al. 2007}
\tablenotetext{i}{AGR09 estimate $m_K\sim10$, Acosta-Pulido et al. 2007 measure $m_K=9.8$ as late as September 2006; thus we adopt 9.9$\pm$0.1 }
\tablenotetext{j}{Most recent published $K-$band measurements is from Aspin et al. 2009. Sept 16, 2008 $m_K=7.53\pm0.09$.}

\end{deluxetable}

\clearpage

\begin{figure} 
\epsscale{0.8}
\plotone{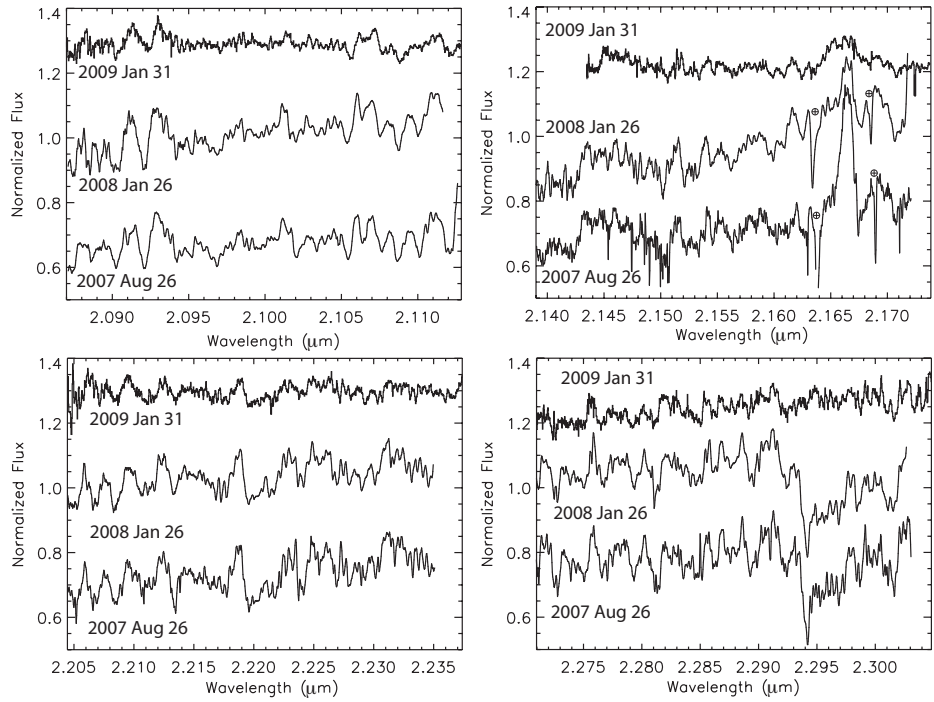}
\caption[]{$K-$band Spectra of V1647 Ori. We compare the $K-$band spectrum we acquired during the 2008 outburst to the quiescent spectra presented by AGR09. The spectrum taken on each date has been offset for clarity. Uncanceled telluric lines are noted with the Earth symbol. Apart from the CO v=2-0 bandhead at 2.294$\micron$ and the Br $\gamma$ line at 2.167$\micron$, the features did not change shape although the veiling of the lines increased. } 
\end{figure}
\clearpage

\begin{figure} 
\epsscale{0.5}
\plotone{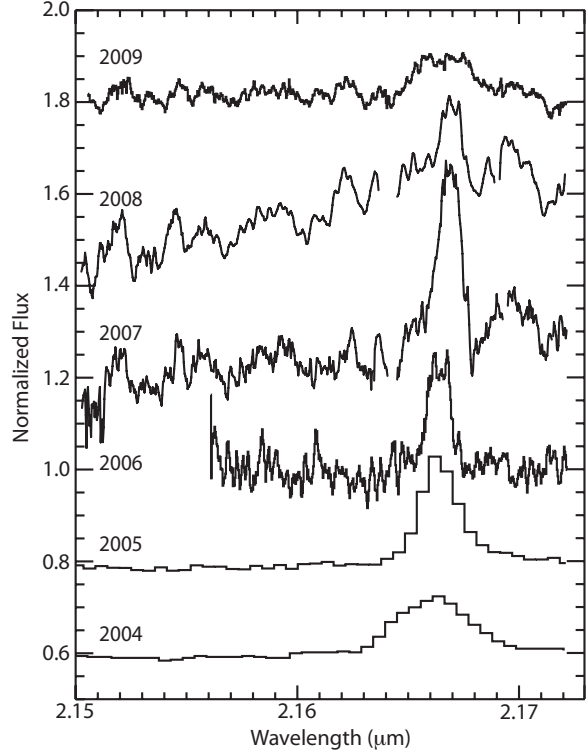}
\caption[]{Evolution of Br $\gamma$ line profile.  We have collected spectra of the Br $\gamma$ line observed annually from 2004-2009 and present them here offset for clarity. At the onset of the 2003 outburst the FWHM of the line was 550 km s$^{-1}$ (Vacca et al. 2004). By 2005, the line had narrowed considerably and the FWHM remained $\sim$150 km s$^{-1}$ during the quiescent phase. At the onset of the 2008 outburst, the line broadened again.}
\end{figure}
\clearpage

\begin{figure} 
\epsscale{1.0}
\plotone{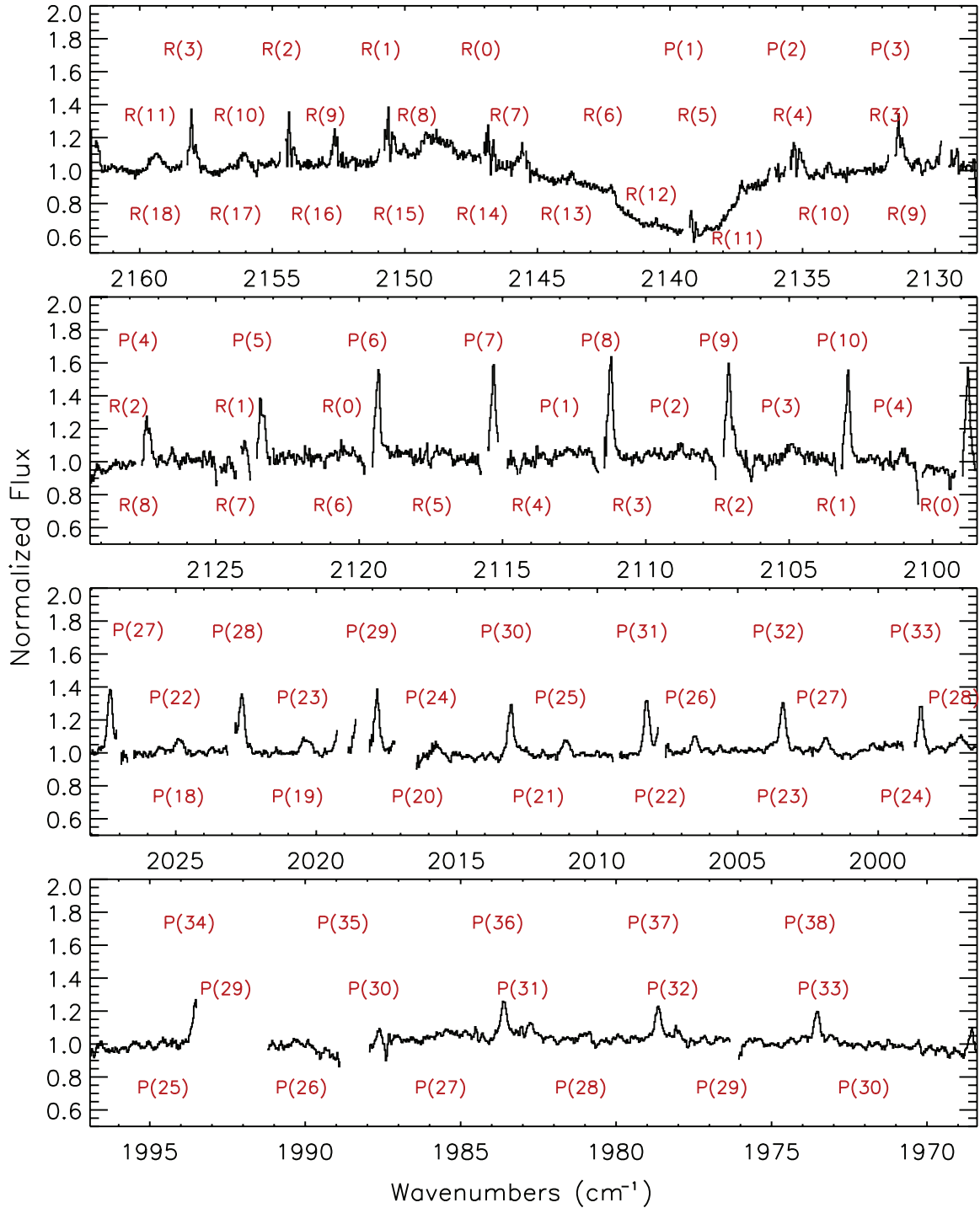}
\caption[]{Fundamental Ro-virbational spectrum of V1647 Ori. The normalized $M-$band spectra of V1647 Ori are labeled such that the line designations of the v=1-0 band are located on the top row, the line designations of the v=2-1 band are located on the middle row and the $^{13}$CO v=1-0 band is labeled on the bottom row. The gaps in the spectrum occur at regions where the atmospheric transmittance is less than 50\%. Superimposed on the low-J lines are narrow absorption features due to interstellar gas. Centered near 2139.5 cm$^{-1}$ is a broad absorption feature due to CO ice.} 
\end{figure}
\clearpage

\begin{figure} 
\epsscale{1.0}
\plotone{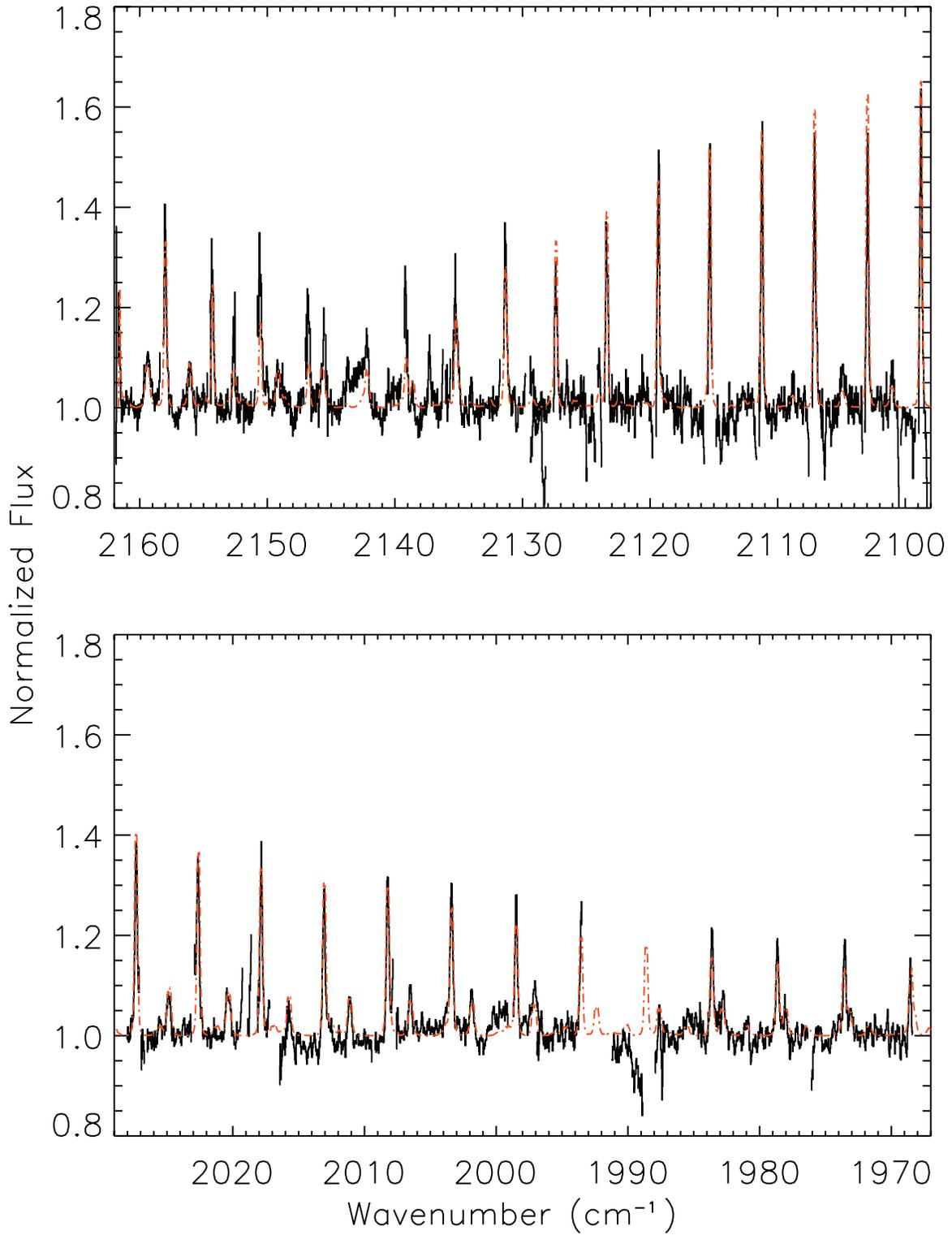}
\caption[]{Fundamental CO ro-vibrational spectrum of V1647 Ori. The ice and foreground gas absorption lines have been ratioed out of the spectrum to facilitate analysis of the circumstellar emission lines. The gaps in the model are regions where the transmittance of the atmosphere is less than 50\%. The red dot-dashed line is the model. The relative strengths of the lines indicate that the gas is $\sim$ 1800K, and the line profile indicate that the gas is distributed from 0.1-12AU. }
\end{figure}
\clearpage

\begin{figure} 
\epsscale{.4}
\plotone{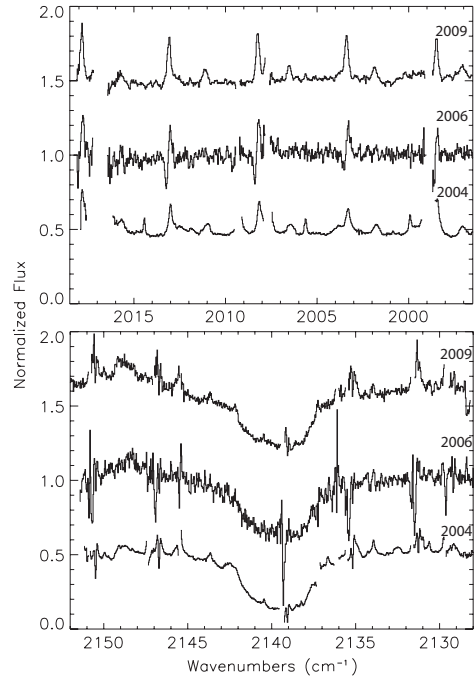}
\caption[]{Evolution of the fundamental CO ro-vibrational spectrum of V1647 Ori. The top panel shows the high-J lines and the lower panel shows the low-J lines observed during three epochs. The spectrum from each epoch has been offset for clarity. The 2004 and 2009 spectra were acquired during the 2003 and 2008 outbursts respectively. The 2006 spectrum was acquired at the onset of the quiescent phase. The CO emission spectrum has evolved considerably between the outbursting and quiescent phases of the star. During the quiescent phase, there appeared to be a short-lived disk wind that led to a warm CO outflow.}
\end{figure}
\clearpage

\begin{figure} 
\epsscale{0.5}
\plotone{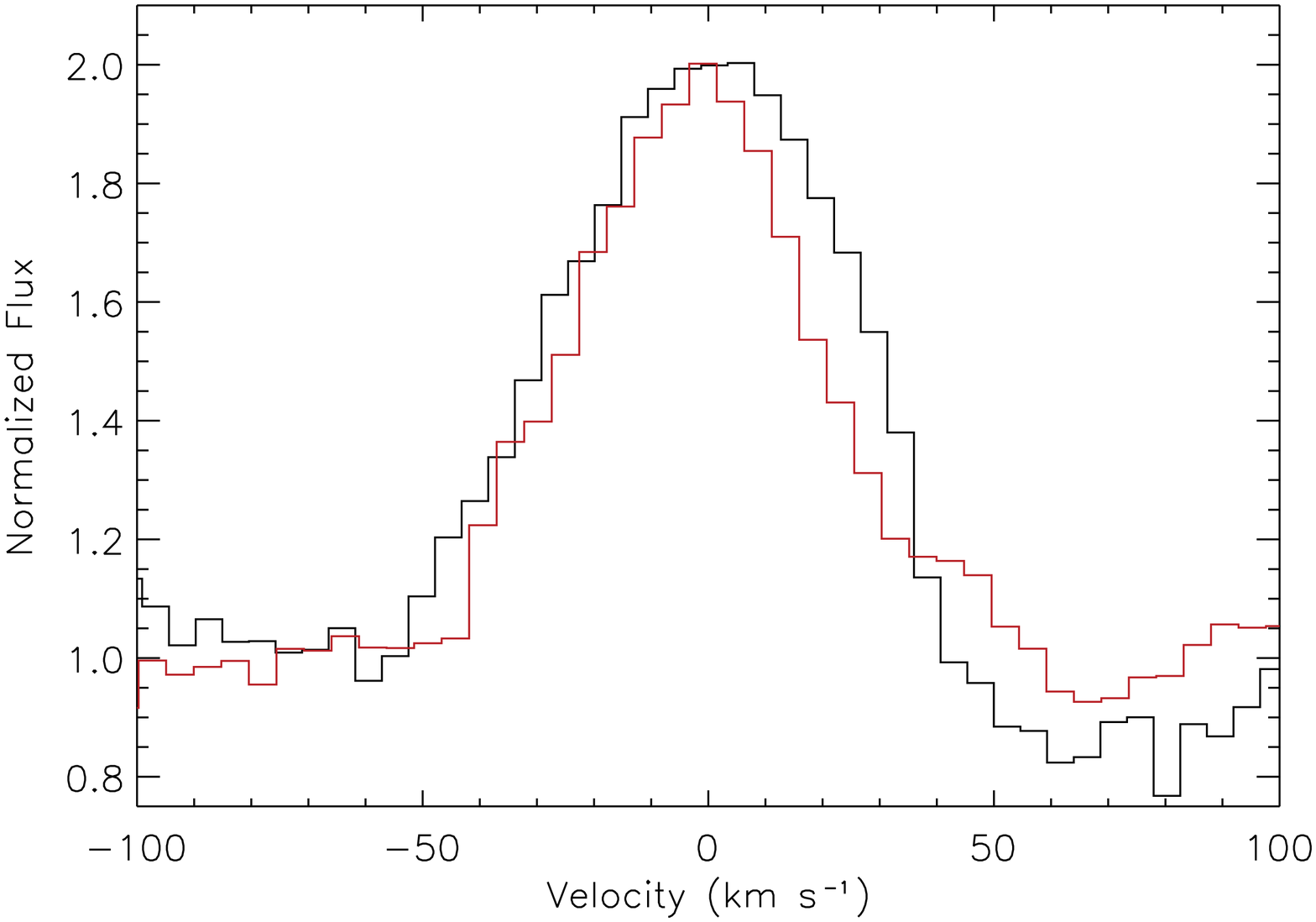}
\caption[]{Comparison of average v=2-1 emission line profile acquired during 2003 (black) and 2008 (red) outbursts. The FWHM of the emission observed during the 2003 outburst is 55 km s$^{-1}$ while the FWHM of the emission observed during the 2008 outburst is 50 km s$^{-1}$. Similarly, the HWZI of the average v=2-1 CO emission line observed during the outburst has decreased from 60 km s$^{-1}$ to 50 km s$^{-1}$.  This suggests that the truncation radius during the 2003 outburst was slightly smaller than during the 2008 outbursts - consistent with the conclusion that the accretion rate during the 2003 outburst was somewhat larger than in the 2008 outburst.}
\end{figure}

\end{document}